\documentclass[conference]{IEEEtran}

\usepackage{subcaption}

\usepackage{cite}
\usepackage{amsmath,amssymb,amsfonts}
\usepackage{graphicx}
\usepackage{textcomp}
\usepackage{xcolor}
\usepackage{color,soul}
\usepackage{multirow}
\usepackage{makecell}
\usepackage{url}
\usepackage{tabularx}
\usepackage{svg}
\usepackage{mathabx}
\usepackage{mathtools}
\usepackage{enumitem}
\usepackage{adjustbox}
\setlist[itemize]{leftmargin=*} \usepackage{algorithm,algorithmic}
\def\BibTeX{{\rm B\kern-.05em{\sc i\kern-.025em b}\kern-.08em
    T\kern-.1667em\lower.7ex\hbox{E}\kern-.125emX}}

\usepackage{float}
\usepackage[hidelinks]{hyperref}
\begin{document}

\title{Exploring the Effects of Load Altering Attacks on Load Frequency Control through Python and RTDS}
\IEEEaftertitletext{\vspace{-10\baselineskip}}

\author{
    \IEEEauthorblockN{
        Michał Forystek\IEEEauthorrefmark{1},
        Andrew D. Syrmakesis\IEEEauthorrefmark{2}, Alkistis Kontou\IEEEauthorrefmark{2}, Panos Kotsampopoulos\IEEEauthorrefmark{2},\\ Nikos D. Hatziargyriou\IEEEauthorrefmark{2}, 
        Charalambos Konstantinou\IEEEauthorrefmark{1}
    }
    
    \IEEEauthorblockA{\IEEEauthorrefmark{1}CEMSE Division, King Abdullah University of Science and Technology (KAUST)}
    \IEEEauthorblockA{\IEEEauthorrefmark{2}School of Electrical and Computer Engineering, National Technical University of Athens}
    \vspace{-10mm}
}

\maketitle

\begin{abstract}
	The modern power grid increasingly depends on advanced information and communication technology (ICT) systems to enhance performance and reliability through real-time monitoring, intelligent control, and bidirectional communication. However, ICT integration also exposes the grid to cyber-threats. Load altering attacks (LAAs), which use botnets of high-wattage devices to manipulate load profiles, are a notable threat to grid stability. While previous research has examined LAAs, their specific impact on load frequency control (LFC), critical for maintaining nominal frequency during load fluctuations, still needs to be explored. Even minor frequency deviations can jeopardize grid operations. This study bridges the gap by analyzing LAA effects on LFC through simulations of static and dynamic scenarios using \texttt{Python} and \texttt{RTDS}. The results highlight LAA impacts on frequency stability and present an eigenvalue-based stability assessment for dynamic LAAs (DLAAs), identifying key parameters influencing grid resilience.
	
\end{abstract}

\vspace{-3mm}
\section{Introduction}
\label{sec:Introduction}
\vspace{-1mm}

The integration of information and communication technology (ICT) has transformed power system operations, enabling real-time monitoring, intelligent control, and bidirectional communication. While these advancements enhance grid efficiency, they also introduce vulnerabilities to cyber-threats  \cite{9351954}. Among these, load altering attacks (LAAs) pose a significant threat by coordinating load changes via botnets of high-wattage devices, disrupting the demand-supply balance and system frequency \cite{BlackIoT, GridShock}. LAAs exploit the relationship between frequency level and the rotational speed of power plant rotors \cite{RobustPowerSystemFrequencyControl}. As  power plants and apparatus are designed to operate within a narrow frequency range, abrupt load changes from LAAs can cause severe frequency fluctuations, risking system instability or damage to its components \cite{BlackIoT, RobustPowerSystemFrequencyControl}.

Due to their critical implications, LAAs have been extensively studied in the literature \cite{surveyLAA}. Static LAAs (SLAAs) are defined as sudden, malicious load alterations that aim to disrupt system stability \cite{originalLAA}. The authors of \cite{BlackIoT, GridShock} examine their effects on the grid, such as frequency imbalance, power line failures, cascading outages, and raised operational costs. While protection mechanisms, such as protective relays and load shedding, help mitigate large-scale blackouts and cascading failures, they cannot entirely prevent local outages or network segmentation caused by SLAAs  \cite{DarkAndGloomy}. In contrast, dynamic LAAs (DLAAs), introduced in \cite{DynamicLAA}, exploit local frequency deviations to amplify instability, exacerbating disruptions. Further variations of DLAAs utilize discrete load manipulations within targeted areas to destabilize the system \cite{DynamicLAADiscrete, DynamicLAADiscrete2}.

To maintain frequency stability, power grids utilize a three-tiered frequency control protection system \cite{RobustPowerSystemFrequencyControl}. When deviations occur, primary control is activated locally at each generator to proportionally counter the deviation by adjusting mechanical input based on reserves. However, primary control alone cannot restore nominal frequency. Secondary or load frequency control (LFC), typically implemented as an integral controller, adjusts generator load reference set points to return the frequency to its target value. If secondary control is insufficient, manual tertiary control addresses the remaining imbalance \cite{RobustPowerSystemFrequencyControl}. Among these, LFC is the key automated mechanism, with its reliability and security extensively studied in the literature \cite{SyrmakesisThesis, DoSOnLFC, DelayedAttackOnLFC, LFCAttacksSummary, 10518173}. For example, \cite{LFCAttacksSummary} reviews cyber-threats to LFC, detailing attack models and mitigation strategies, while \cite{SyrmakesisThesis, 10518173} analyze LFC vulnerabilities from ICT integration, emphasizing the need for robust protective measures.

Recognizing the vital role of LFC in grid stability and the impacts of LAAs, recent studies have investigated their intersection. In \cite{LAALFCTolerantFramework}, the authors introduce a model-free LFC framework to defend against LAAs using active and passive strategies. The active strategy enables LFC to learn from LAA patterns, improving its ability to mitigate future attacks. In contrast, the passive strategy employs reinforcement learning to adapt defense policies dynamically with each attack. Further evaluation of this framework against a traditional controller under LAA and DoS attacks is conducted in \cite{LAALFCModelFree2}, demonstrating its effectiveness through simulation. 
While extensive research exists on LAAs and LFC, as well as on their interaction, the specific impact of LAA on LFC dynamics and system stability still needs to be explored. To address this gap, our paper contributes the following:
\vspace{-1mm}
\begin{itemize}
	\item We conduct \texttt{Python} and \texttt{RTDS} simulations of LFC under SLAAs and DLAAs on the IEEE 39-bus system, capturing theoretical and real-time dynamics to analyze LFC.
	\item In multiple scenarios, we systematically evaluate the responses of LFC, focusing mainly on how LFC manages frequency deviations and load fluctuations induced by LAAs.
	\item We conduct a system stability assessment under DLAA conditions, utilizing eigenvalue analysis to explore the effect of different parameter variations.
\end{itemize}

\vspace{-1mm}
The paper is organized as follows. Section II introduces the models of LFC and LAA. Section III describes the implementation setup, defines simulated scenarios, and discusses the results. Finally, Section IV summarizes the article.

\vspace{-1mm}
\section{Analytical Models of LFC and LAAs}
\label{sec:ModelMethodology}

This section introduces the analytical models of LFC, SLAA, and DLAA, formulated in state-space representation, enabling efficient software implementation. The LFC model is also depicted through a block diagram of transfer functions, providing a clear visualization of control dynamics. Finally, we integrate LAA into the LFC model, allowing for a comprehensive analysis of its impact on frequency control.

\vspace{-1mm}
\subsection{LFC Modeling}
\label{subsec:ModelOfLoadFrequencyControl}

\vspace{-1mm}
First, we introduce the analytical model of LFC, which considers the $N$-area system as defined in \cite{RobustPowerSystemFrequencyControl}. Each area is individually modeled while accounting for inputs from directly connected neighboring areas. This model encompasses frequency deviation, primary and secondary control mechanisms, and dynamic load adjustments—an essential feature enabling the effective simulation and analysis of LAAs. The model can be represented as state-space equations, more suitable for discrete-time implementation in computational environments, or as a block diagram of transfer functions, providing a clear visual interpretation of control interactions.

\vspace{-2mm}
\small
\begin{equation}
	\label{eq:stateSpace}
	\dot{x_i} = A_ix_i + B_{1i}w_i + B_{2i}u_i;\qquad
	y_i = C_ix_i
\end{equation}
\normalsize
\vspace{-5mm}

Starting with the state-space representation shown in \eqref{eq:stateSpace}, the internal state of the $i$-th area is captured as a vector $x_i$ \eqref{eq:stateVector} comprising four components. The $\Delta f_i$ is the area's frequency deviation. The $\Delta P_{tie, i}$ is a total tie-line power change between the area and all other directly connected areas. Finally, the $x_{mi}$ \eqref{eq:xmi} and $x_{gi}$ \eqref{eq:xgi}, modeled as vectors to consider the presence of $n$ generators in one area, contain outputs of all turbines $\Delta P_m$ (mechanical power change), and all governors $\Delta P_g$ (valve position change) in the area, respectively.

\vspace{-2mm}
\small
\begin{align}
	\label{eq:stateVector}
	x_i^T &= \begin{bmatrix}
		\Delta f_i & \Delta P_{tie,i} & x_{mi} & x_{gi}
	\end{bmatrix} \\
	\label{eq:xmi}
	x_{mi}^T &= \begin{bmatrix}
		\Delta P_{m1i} & \Delta P_{m2i} & \cdots & \Delta P_{mni}
	\end{bmatrix} \\
	\label{eq:xgi}
	x_{gi}^T &= \begin{bmatrix}
		\Delta P_{g1i} & \Delta P_{g2i} & \cdots & \Delta P_{gni}
	\end{bmatrix}
\end{align}
\normalsize
\vspace{-4mm}

Vector $w_i$ \eqref{eq:inputVector1AndInputFromOtherAreas} represents the input to the area as an area's load change $\Delta P_{Li}$ and the input from all directly connected areas $v_i$. The $v_i$ \eqref{eq:inputVector1AndInputFromOtherAreas} is a sum of the tie-lines synchronizing torque coefficients $T_{ij}$ between area $i$ and each directly connected area $j$, times $\Delta f_j$ of those areas. In $T_{ij}$, defined in \eqref{eq:torqueCoefficientAndInputVector2}, $V$ is the area's voltage at the equivalent machine's terminals, $X_{ij}$ is the tie-line reactance, and $(\delta_i^0$, $\delta_j^0)$ is the equilibrium point of voltage angles $\delta_i$ and $\delta_j$ of the areas' equivalent machines. The second input vector $u_i$ \eqref{eq:torqueCoefficientAndInputVector2} represents the area's LFC controller output with the function $K_i(\cdot)$ identifying its dynamics.

\vspace{-4mm}
\small
\begin{align}
	\label{eq:inputVector1AndInputFromOtherAreas}
	w_i^T &= \begin{bmatrix}
		\Delta P_{Li} & v_i
	\end{bmatrix};
	&v_i &= \sum_{\mathclap{\substack{j=1\:j\neq i}}}^{N}T_{ij}\Delta f_j \\
	\label{eq:torqueCoefficientAndInputVector2}
	T_{ij} &= \frac{|V_i||V_j|}{X_{ij}}cos(\delta^0_i-\delta^0_j);
	&u_i &= \Delta P_{Ci} = K_i(ACE_i)
\end{align}
\normalsize
\vspace{-3mm}

The system's output $y_i$ \eqref{eq:outputVector} acts as the LFC controller input called the area control error (ACE). In ACE, the $\beta_i$ \eqref{eq:betaANDRSys} is the bias factor computed by adding the equivalent load damping coefficient $D_i$ and reciprocal of $R_{sys, i}$. The $R_{sys, i}$ is defined in \eqref{eq:betaANDRSys} with $R_{ki}$ being the generator's droop characteristic.

\vspace{-2mm}
\small
\begin{equation}
	\label{eq:outputVector}
	y_i = ACE_i = \beta_i\Delta f_i + \Delta P_{tie,i}
\end{equation}
\begin{equation}
	\label{eq:betaANDRSys}
	\beta_i = D_i + \frac{1}{R_{sys,i}};\qquad
	\frac{1}{R_{sys,i}} = \sum_{\substack{k=1}}^{n}\frac{1}{R_{ki}} 
\end{equation}
\normalsize
\vspace{-2mm}

The system matrix $A$ is defined in \eqref{eq:matrixAndA11} to \eqref{eq:matrixA31And32And33}. $H_i$ is an equivalent generator inertia constant. $T_t$ and $T_G$ are turbine and governor time constants, respectively.

\vspace{-3mm}
\small
\begin{align}
	\label{eq:matrixAndA11}
	A_i& = \begin{bmatrix}
		A_{i11} & A_{i12} & A_{i13} \\
		A_{i21} & A_{i22} & A_{i23} \\
		A_{i31} & A_{i32} & A_{i33}
	\end{bmatrix};\enspace
	A_{i11} = \begin{bmatrix}
		\frac{-D_i}{2H_i} & \frac{-1}{2H_i} \\
		\displaystyle 2\pi\sum_{\mathclap{\substack{j=1\ j\neq i}}}^{N}T_{ij} & 0
	\end{bmatrix} \\
	\label{eq:matrixA12AndA13And21}
	A_{i12}& = \begin{bmatrix}
		\frac{1}{2H_i} & \kern-0.7em ...\kern-0.7em & \frac{1}{2H_i} \\
		0 & \kern-0.7em ...\kern-0.7em & 0
	\end{bmatrix}_{2\times n};\quad\!\:
	A_{i13} = 0_{2\times n};\:\:
	A_{i21} = 0_{n\times 2} \\
	\label{eq:matrixA22AndA23}
	A_{i22}& = diag\begin{bmatrix}
		\frac{-1}{T_{t1i}} & \kern-0.7em ...\kern-0.7em & \frac{-1}{T_{tni}}
	\end{bmatrix};\:\:
	A_{i23} = diag\begin{bmatrix}
		\frac{1}{T_{t1i}} & \kern-0.7em ...\kern-0.7em & \frac{1}{T_{tni}}
	\end{bmatrix} \\
	\label{eq:matrixA31And32And33}
	A_{i31}& = \begin{bmatrix}
		\!\frac{-1}{T_{g1i}R_{1i}} & \kern-0.6em 0 \\
		\!\vdots & \kern-0.6em\vdots \\
		\!\frac{-1}{T_{gni}R_{ni}} & \kern-0.6em 0
	\end{bmatrix}\!;\:
	A_{i32} = 0_{n\times n};\:
	A_{i33} = diag\begin{bmatrix}
		\!\frac{-1}{T_{g1i}\!} \\
		\!\vdots\! \\
		\!\frac{-1}{T_{gni}\!}
	\end{bmatrix}^T
\end{align}
\normalsize
\vspace{-2mm}

The model divides the system input into two parts: \textit{(i)} the uncontrollable load changes inside the area and the inputs from the other areas, and \textit{(ii)} the controllable LFC controller response. Thus, two input matrices, $B_1$ \eqref{eq:matrixB1AndB11AndB12AndB13} and $B_2$ \eqref{eq:matrixB2AndB21AndB22AndB23} correspond to input vectors $w$ and $u$. The $\alpha_k$ is the LFC participation factor of the area's $k$-th generator. The sum of all $\alpha$ from the same area must equal one. If the generator does not participate in the LFC, $\alpha$ equals zero. This parameter is time-dependent and should be computed dynamically \cite{RobustPowerSystemFrequencyControl}. Finally, the output matrix $C$ is defined in \eqref{eq:matrixC}.

\vspace{-4mm}
\small
\begin{align}
	\label{eq:matrixB1AndB11AndB12AndB13}
	B_{1i} &= \begin{bmatrix}
		B_{1i1} \\
		B_{1i2} \\
		B_{1i3}
	\end{bmatrix};\quad
	B_{1i1} = \begin{bmatrix}
		\frac{-1}{2H_i} & 0 \\
		0 & -2\pi
	\end{bmatrix};\quad
	\begin{aligned}
		B_{1i2} = 0_{n\times 2} \\
		B_{1i3} = 0_{n\times 2}
	\end{aligned} \\
	\label{eq:matrixB2AndB21AndB22AndB23}
	B_{2i} &= \begin{bmatrix}
		B_{2i1} \\
		B_{2i2} \\
		B_{2i3}
	\end{bmatrix};\quad
	\begin{aligned}
		B_{2i1} = 0_{2\times 1} \\
		B_{2i2} = 0_{n\times 1}
	\end{aligned}\:;\quad
	B_{2i3} = \begin{bmatrix}
		\frac{\alpha_{1i}}{T_{g1i}} &
		\kern-0.7em ...\kern-0.7em &
		\frac{\alpha_{ni}}{T_{gni}} 
	\end{bmatrix}
\end{align}
\vspace{-2mm}
\vspace{-2mm}
\begin{equation}
	\label{eq:matrixC}
	C_{i} = \begin{bmatrix}
		\beta_i & 1 & 0_{1\times n} & 0_{1\times n}
	\end{bmatrix}
\end{equation}
\normalsize
\vspace{-4mm}

\begin{figure}[t]
	\centering
	\includegraphics[width=0.9\linewidth]{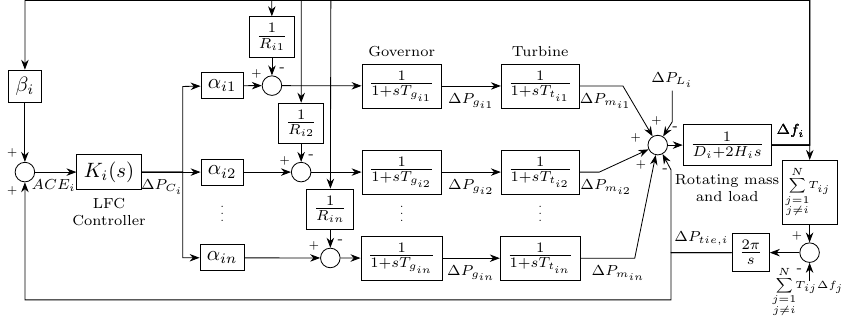}
	\vspace{-2mm}
	\caption{Block diagram of the LFC model $i^{th}$ area with $k$ generators.}
	\vspace{-5mm}
	\label{fig:area_schema_full}
\end{figure}

Another way to visualize the system is through the block diagram of the area, as illustrated in Fig.  \ref{fig:area_schema_full}. It employs transfer functions to represent different segments of the system. In this diagram, the upper portion depicts the primary control for each generator, achieved by multiplying $\Delta f$ by the reciprocal of $R$. It responds locally to frequency changes, adjusting the generator's output. Meanwhile, the secondary control loop, shown as the leftward input to the governor, adjusts the area-wide load reference setpoint to restore nominal frequency. While primary control parameters are specific to each generator, the LFC operates at the area level, distributing corrective actions among participating generators according to their respective $\alpha$. Detailed state-space and block diagram model definitions are available in \cite{RobustPowerSystemFrequencyControl}.

\vspace{-1mm}
\subsection{LAA Modeling}
\label{subsec:ModelOfStaticAndDynamicLAA}
\vspace{-1mm}

SLAAs can be incorporated into the LFC model as they only affect $\Delta P_L$. This additively affects $\Delta f$ by $\frac{-\Delta P_L}{2H}$, influencing all the other variables dependent on such deviation. Alternatively, SLAAs can be modeled as a set of differential equations \cite{lakshminarayana2022load}. In this case, as shown in \eqref{eq:staticLAA}, the system's internal state is modeled as a concatenation of vectors $\delta$, $\theta$, and $\omega$: voltage phase angles of generator buses,  voltage phase angles of load buses, and rotor frequency deviation of generator buses. The vector $P^{LS}$ defines the system input as a secure portion of the load at each bus. In contrast, vector $\epsilon^L$ represents the vulnerable portion of the load used to perform SLAA.

\small
\vspace{-1mm}
\begin{equation}
	\label{eq:staticLAA}
	E \begin{bmatrix}
		\dot{\delta} \\ 
		\dot{\theta} \\
		\dot{\omega}
	\end{bmatrix} = A \begin{bmatrix}
		\delta \\
		\theta \\
		\omega
	\end{bmatrix} + B(P^{LS} + \epsilon^L)
\end{equation}
\vspace{-2.5mm}
\normalsize

The \eqref{eq:staticLAAMatrixAAndMatrixB}, and \eqref{eq:imagAdmittanceMatrix} show the system, input, and mass matrices $A$, $B$, and $E$. We define $A^{\kern-0.1em 1} \!\!:=\! diag(\!A\!*\!\mathbf{1}\!)$ where $\mathbf{1}$ is the column vector of ones. The diagonal matrices $M$, $D^G$, $K^P$, and $K^I$ show the generator inertias, damping coefficients, and proportional and integral coefficients for primary and secondary control accordingly. The admittance matrix $H_{bus}$ \eqref{eq:imagAdmittanceMatrix} depicts connections between generator-to-generator ($H^{GG}$), generator-to-load bus ($H^{GL}$), load bus-to-generator ($H^{LG}$), and load bus-to-load bus ($H^{LL}$). If two buses are unconnected, the respective matrix element equals zero. Lastly, $I$ is the identity matrix of the appropriate dimensions.

\small
\vspace{-3mm}
\begin{equation}
	\label{eq:staticLAAMatrixAAndMatrixB}
	\setlength{\arraycolsep}{2pt}
	A = \begin{bmatrix}
		0 & 0 & I \\
		-\!H^{\!LG} & H^{\!LG^1}\kern-0.6em+\!\!H^{\!LL^1}\kern-0.7em-\!\!H^{\!LL} & 0 \\
		K^{\!I}\kern-0.4em+\!\!H^{\!GG^1}\kern-0.7em-\!\!H^{\!GG}\kern-0.4em+\!\!H^{\!GL^1} & -\!H^{\!GL} & K^{\!P}\kern-0.5em+\!\!D^{\kern-0.07emG} 
	\end{bmatrix}
\end{equation}
\vspace{-2mm}
\begin{equation}
	\label{eq:imagAdmittanceMatrix}
	E = \begin{bmatrix}
		I & 0 & \kern-0.5em 0 \\
		0 & 0 & \kern-0.5em 0 \\
		0 & 0 & \kern-0.5em -M
	\end{bmatrix};\quad
	B = \begin{bmatrix}
		0 \\
		I \\
		0
	\end{bmatrix}; \quad
	H_{bus} = \begin{bmatrix}
		H^{GG} H^{GL} \\
		H^{LG} H^{LL}
	\end{bmatrix}
\end{equation}
\vspace{-2mm}
\normalsize

While the previous model effectively represents SLAA, this attack does not alter the system stability, a key distinction of DLAA \cite{DynamicLAA}. To model DLAA, first we modify \eqref{eq:staticLAA} by substituting $\theta$ \eqref{eq:theta} to then obtain the non-descriptor form. Next, we incorporate the attack into the system matrix to affect stability. As shown in \eqref{eq:stateSpaceDynamic}, the DLAA proportional coefficient vector $K^{LG}>\nolinebreak0$ indicates manipulation of the vulnerable load portion that can modify the system state during an attack.

\small
\vspace{-3mm}
\begin{equation}
	\label{eq:theta}
	\theta = H^{inv}(H^{LG}\delta\! - \!P^L); \quad H^{inv}\kern-0.2em = (H^{LG^1}\kern-0.4em+\!H^{LL^1}\kern-0.5em-\!H^{LL})^{-1}
\end{equation}
\begin{equation}
	\label{eq:stateSpaceDynamic}
	\begin{bmatrix}
		\dot{\delta} \\ \dot{\omega}
	\end{bmatrix} = A'\begin{bmatrix}
		\delta \\ \omega
	\end{bmatrix} + B'\left(\begin{bmatrix}
		0 & \kern-0.4em -K^{LG}
	\end{bmatrix}\begin{bmatrix}
		\delta \\ \omega
	\end{bmatrix}+P^{LS}\right)
\end{equation}
\vspace{-2mm}
\normalsize

The system matrix also changes as in \eqref{eq:matrixADynamicNew}. We obtain the new system matrix $A^*$ \eqref{eq:matrixA*Dynamic} of the system under attack. The new form of the system state-space equations is shown in \eqref{eq:stateSpaceDLAAFinal}. This model integrates the DLAA into the system matrix, meaning that changes to $K^{LG}$ can affect system stability by shifting the eigenvalues of $A^*$.

\small
\vspace{-1.5mm}
\begin{equation}
	\label{eq:matrixADynamicNew}
	A^* = A' + B'\begin{bmatrix}
		0 & \kern-0.4em -K^{LG}
	\end{bmatrix}
\end{equation}

\vspace{-3mm}
\begin{align}
	\label{eq:matrixA*Dynamic}
	&\begin{aligned}
		A^*\! &=\! \left[\begin{matrix}
			0 \\
			M^{-1}(H^{GG}\kern-0.4em-\!\!H^{GG^1}\kern-0.6em-\!\!H^{GL^1}\kern-0.5em+\!\!H^{GL}H^{inv}H^{\!LG}\kern-0.3em-\!\!K^I)
		\end{matrix}\right.\\
		&\hspace{8.2em}
		\left.\begin{matrix}
			I \\
			-M^{-1}(K^P\kern-0.4em+\!D^G\kern-0.3em+\!H^{GL}H^{inv}K^{LG})
		\end{matrix}\right]
	\end{aligned}
\end{align}

\begin{equation}
	\label{eq:stateSpaceDLAAFinal}
	\begin{bmatrix}
		\dot{\delta} \\ \dot{\omega}
	\end{bmatrix} = A^*\begin{bmatrix}
		\delta \\ \omega
	\end{bmatrix} +\begin{bmatrix}
		0 \\ M^{-1}H^{GL}H^{inv}
	\end{bmatrix}P^{LS}
\end{equation}
\normalsize

\vspace{-2mm}
\subsection{LAA effect on LFC}
\label{subsec:ConncetionOfModelsOfLFCAndLAA}
\vspace{-1mm}

In the context of LFC, we consider the general form of LAA from a power flow perspective \cite{LAALFCTolerantFramework}, as shown in \eqref{eq:generalLAA}. The $P$ represents the original power, and $\mathcal{N}_e$ denotes the set of buses with vulnerable loads indexed by $i$. Buses directly connected to bus $i$ are indexed by $j$, $U$ represents the voltage magnitude at each bus, and $\theta_{ij}$ is the phase angle difference between buses $i$ and $j$. The $G_{ij}$ and $B_{ij}$ are the real and imaginary parts of the admittance between buses $i$ and $j$, respectively, and $d$ is the load manipulation introduced by the LAA.

\vspace{-4mm}
\small
\begin{equation}
	\label{eq:generalLAA}
	P_{is} + d = U_i \sum_{j\in\mathcal{N}_i} U_j (G_{ij}cos(\theta_{ij}) + B_{ij}sin(\theta_{ij}), \forall i \in \mathcal{N}_e
\end{equation}
\normalsize
\vspace{-3mm}

When modeling the LFC under LAA, the general state-space representation can be expressed by \eqref{eq:generalLAALFC}. However, the power flow equations must adhere to the form in \eqref{eq:generalLAA}. In alignment with the previous LFC model, $x$ is the system state as defined in \eqref{eq:stateVector} and $y$ is the system output found in \eqref{eq:outputVector}. For clarity, $u$ is represented in \eqref{eq:inputVector1AndInputFromOtherAreas} and \eqref{eq:torqueCoefficientAndInputVector2} as separate vectors. The $f$ and $g$ are algebraic functions, while $d$ represents the LAA alterations. For an SLAA, $d$ affects the input vector $u$, while for a DLAA, $d$ influences the system matrix incorporated within $f$.

\vspace{-2mm}
\small
\begin{equation}
	\label{eq:generalLAALFC}
	\dot{x} = f(x, u, d);\qquad
	y = g(x)
\end{equation}
\normalsize
\vspace{-3mm}

\vspace{-4mm}
\section{Implementation and simulation results}
\label{sec:ImplementationAndSimulationResutls}

With the models defined, we proceed to examine the impact of LAA on LFC and system stability through simulations. This section details their setup and implementation using \texttt{Python} and \texttt{RTDS} platforms to capture theoretical and real-time dynamics, respectively. We analyze several attack scenarios, observing how LAAs influence frequency response and stability. Results indicate that while \texttt{Python} simulations offer insight into unconstrained system response, the \texttt{RTDS} simulations capture the effects of system limitations, such as finite power reserves and generator non-linearities, on stability. Finally, we present an eigenvalue-based stability analysis under DLAA, highlighting parameter sensitivities critical for grid resilience.

\vspace{-2mm}
\subsection{Simulation Setup}
\label{subsec:Setup}
\vspace{-1mm}

We developed two simulations based on the IEEE 39-bus benchmark, dividing the system into three areas to define the tie-lines required for the LFC algorithm. The first simulation, implemented in \texttt{Python} and available as an open-source project at \cite{PythonSimulationGithub}, explores the impact of SLAAs on LFC, directly implementing the models described in Section \ref{sec:ModelMethodology}. The second simulation implements LFC in \texttt{RSCAD/RTDS}  and examines the role of LAA within the IEEE 39-bus system \cite{RTDSWebsite}. Additionally, we used \texttt{MATLAB} to model DLAA as described in   \ref{subsec:ModelOfStaticAndDynamicLAA} to assess system stability under this attack.

In the \texttt{Python} simulation, we first compute the system matrices $A$, $B$, and $C$. Since the model in  \ref{subsec:ModelOfLoadFrequencyControl} is in continuous time, we convert these matrices to discrete time using \verb|cont2discrete()| with a zero-order hold from the \verb|scipy.signal| library for efficient implementation. We then define a PID-based LFC controller with manually tuned coefficients as the first input and load changes over time as the second input, initializing all system conditions to zero.

The \texttt{RTDS} simulation tests the DLAA's impact on LFC and previously simulated scenarios to compare analytical and real-time results. We implemented the LFC controller on one generator per area, where the generator’s angular frequency is measured, converted to a per-unit deviation from nominal, multiplied by $\beta$, and adjusted for tie-line inputs before passing through an integral controller. A manually derived constant compensates for errors, setting the governor's load reference point. Static and multistep LAAs are implemented with \textit{dynamic load}  \texttt{RSCAD} components. For DLAA, we added proportional controllers at selected loads to exacerbate frequency deviations by applying an additional load based on the proportional coefficient $K^{LG}$ and the base load.

\vspace{-1mm}
\subsection{Scenarios}
\label{subsec:SimulationScenarios}
\vspace{-1mm}

To evaluate the impact of LAA on LFC, we simulated three attack scenarios. In each scenario, the attack begins $30$ seconds after the simulation starts to exclude any initial setup disturbances. In \textbf{Scenario I}, we simulate five variations: a 10\% load increase concentrated in a single area (\textbf{I.1}) and distributed load increases of 10\%, 20\%, and 50\% across multiple areas (\textbf{I.2}, \textbf{I.3}, and \textbf{I.4}, respectively). Scenario \textbf{I.5} applies the maximum load increase across all areas for which the system still manages to restore nominal frequency, with an increase of 16\% in both simulations. \textbf{Scenario II} is an incrementally distributed over time version of I.5. Here, the increases are set at 16\% and 17\% for the \texttt{Python} and \texttt{RTDS} simulations, respectively. \textbf{Scenario III} tests a DLAA by targeting two load buses, simulating the effects of a coordinated, dynamic load alteration on system stability.

\vspace{-2mm}
\subsection{Results}
\label{subsec:SimulationResults}
\vspace{-1mm}

The frequency plots generated in both simulations include three pairs of lines indicating the thresholds for plant and apparatus operation requirements, as specified in the Saudi Arabia Grid Code \cite{saudiGrid}. The generator plants and apparatus are designed to operate within a frequency range of $57.0$ Hz to $62.5$ Hz. We consider an attack successful if the frequency exceeds this range or maintains it within defined thresholds for longer than the specified operational limits.

\begin{figure*}
	\begin{subfigure}[t]{0.19\textwidth}
		\centering
		\def\svgwidth{\linewidth}
		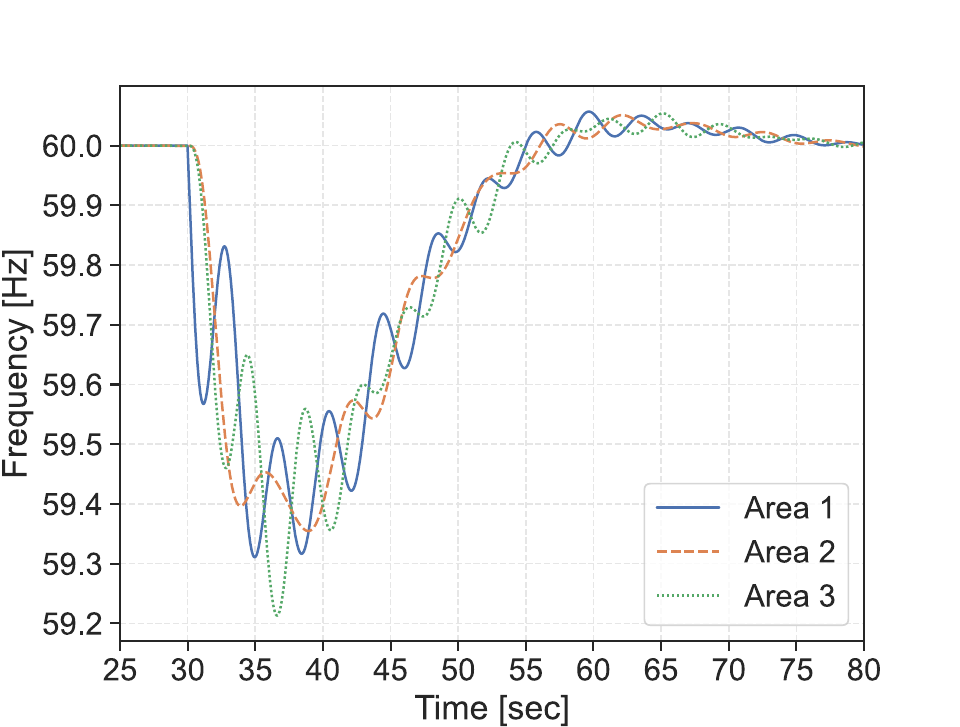
		\vspace{-6mm}
		\caption{\texttt{Python} scenario I.1}
		\label{fig:pySimple10upArea1}
	\end{subfigure}
	\begin{subfigure}[t]{0.19\textwidth}
		\centering
		\def\svgwidth{\linewidth}
		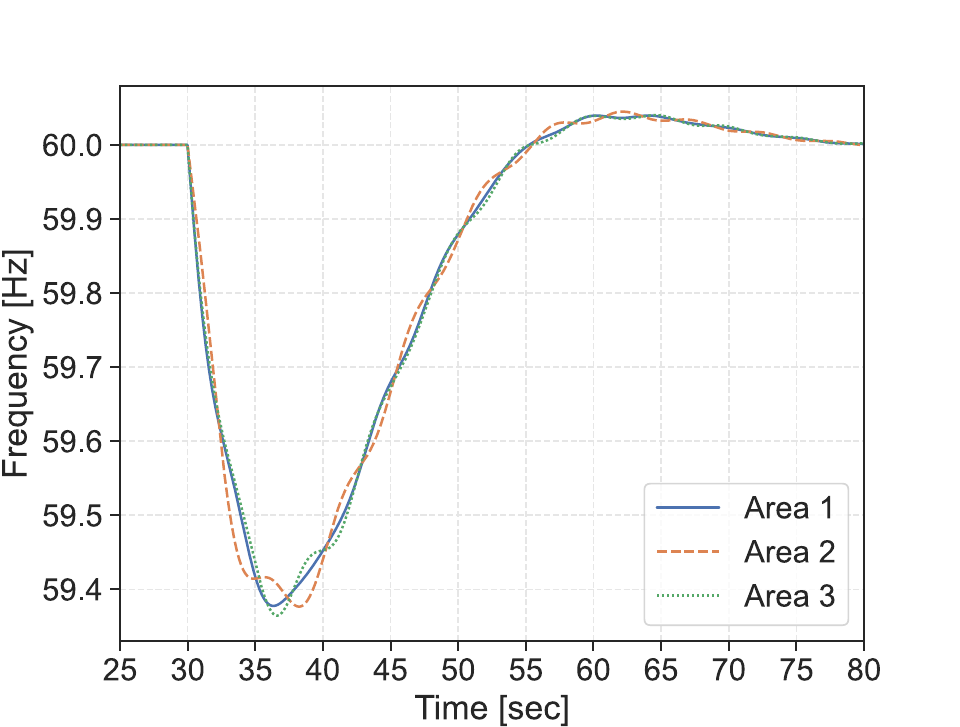
		\vspace{-6mm}
		\caption{\texttt{Python} scenario I.2}        \label{fig:pySimple10upAll}
	\end{subfigure}
	\begin{subfigure}[t]{0.19\textwidth}
		\centering
		\def\svgwidth{\linewidth}
		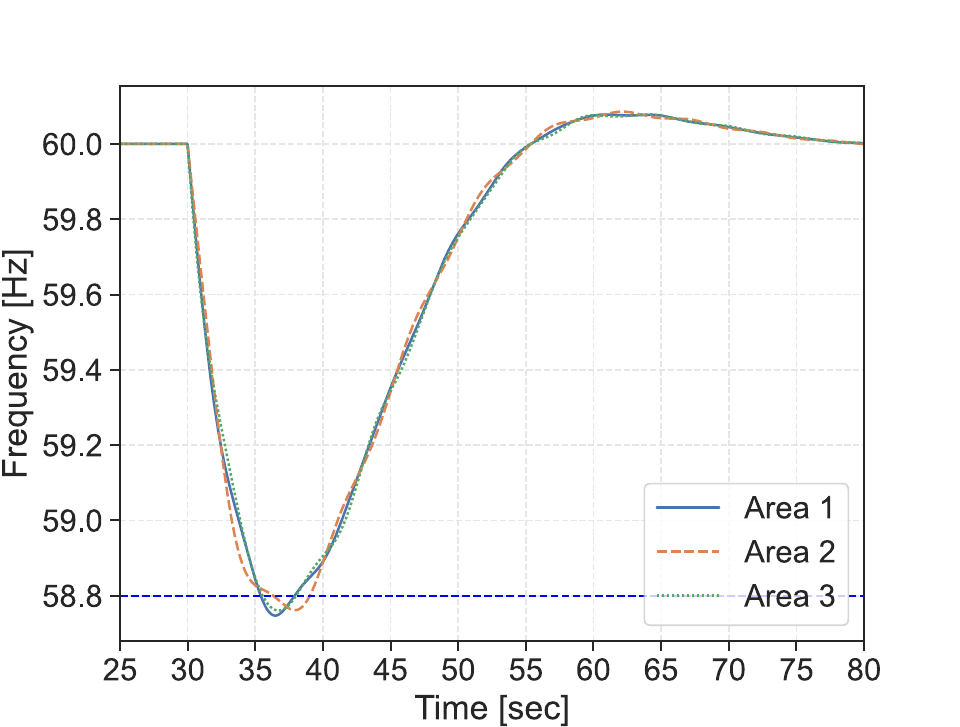
		\vspace{-6mm}
		\caption{\texttt{Python} scenario I.3}        \label{fig:pySimple20upAll}
	\end{subfigure}
	\begin{subfigure}[t]{0.19\textwidth}
		\centering
		\def\svgwidth{\linewidth}
		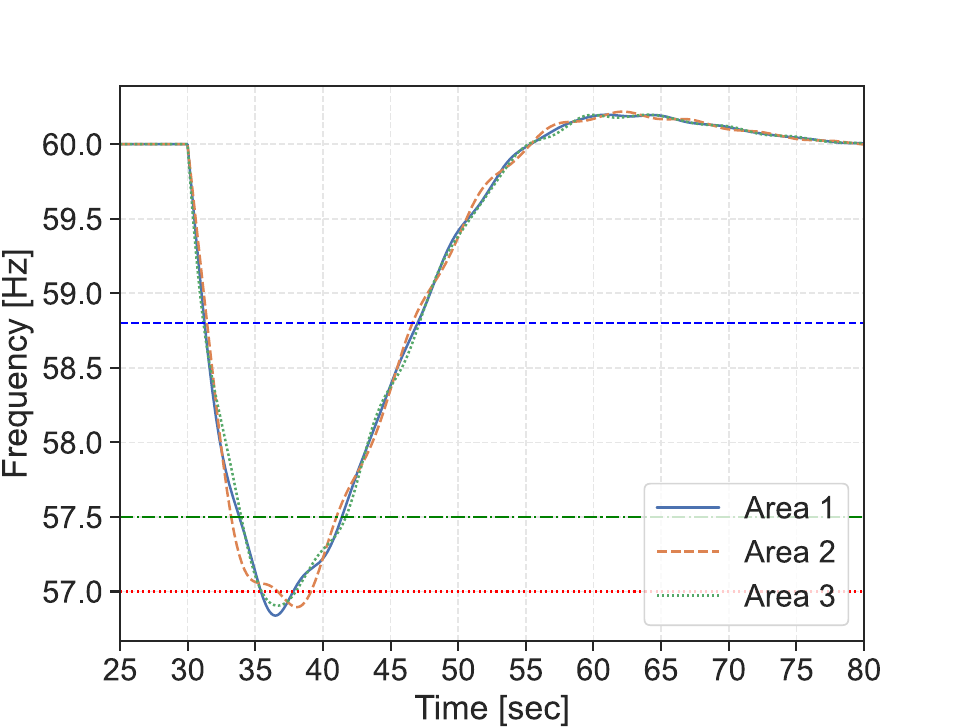
		\vspace{-6mm}
		\caption{\texttt{Python} scenario I.4}        \label{fig:pySimple50upAll}
	\end{subfigure}
	\begin{subfigure}[t]{0.19\textwidth}
		\centering
		\def\svgwidth{\linewidth}
		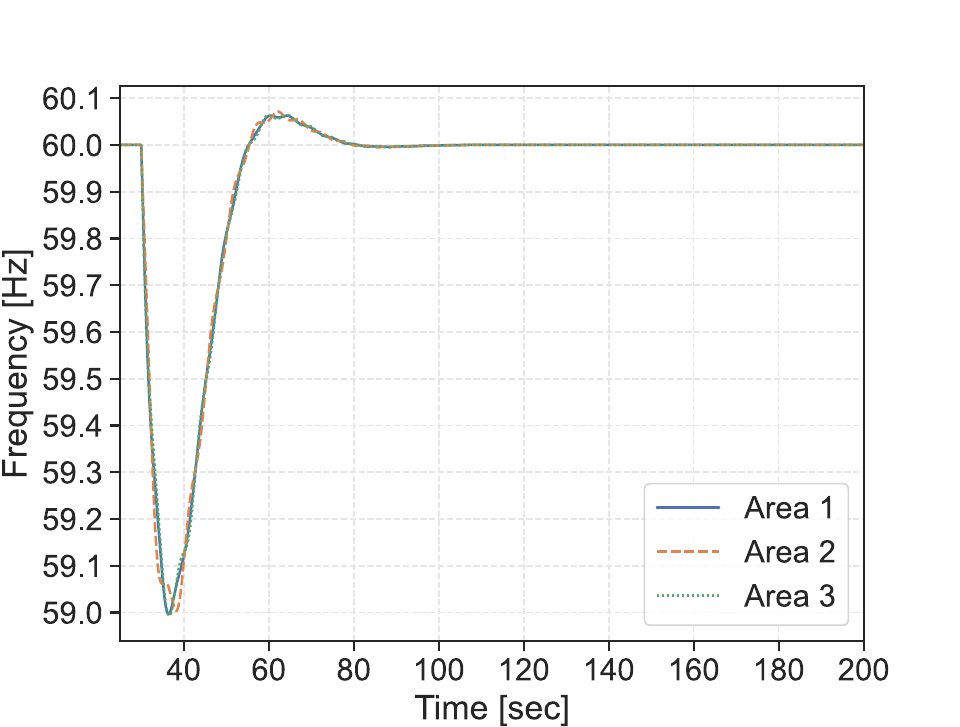
		\vspace{-6mm}
		\caption{\texttt{Python} scenario I.5}        \label{fig:pyMaxInAllAreas}
	\end{subfigure}
	
	\hfill
	
	\begin{subfigure}[t]{0.19\textwidth}
		\centering
		\includegraphics[width=0.9\linewidth]{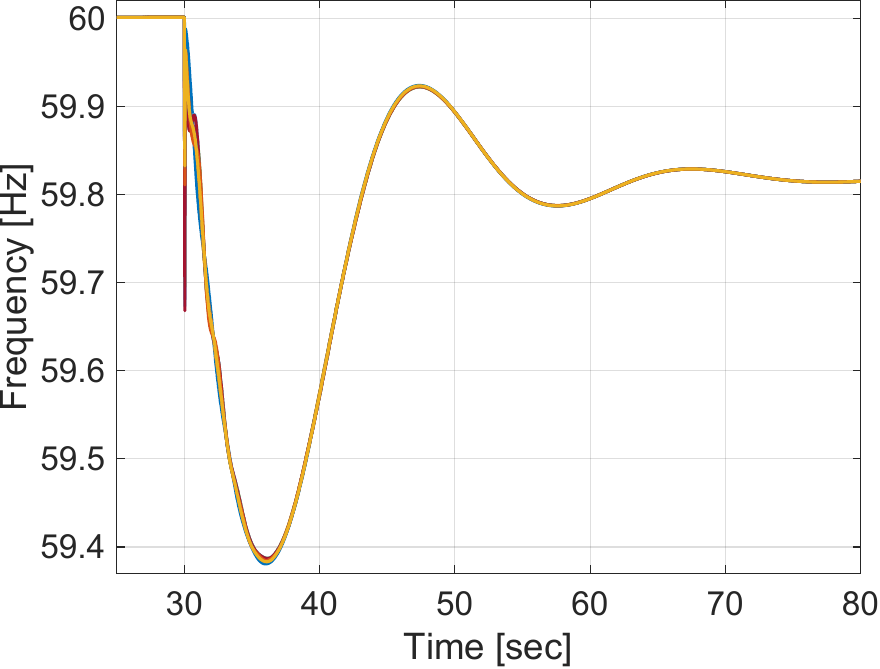}
		\vspace{-2mm}
		\caption{\texttt{RTDS} scenario I.1}
		\label{fig:LFC10increaseSingleArea}
	\end{subfigure}
	\begin{subfigure}[t]{0.19\textwidth}
		\centering
		\includegraphics[width=0.9\linewidth]{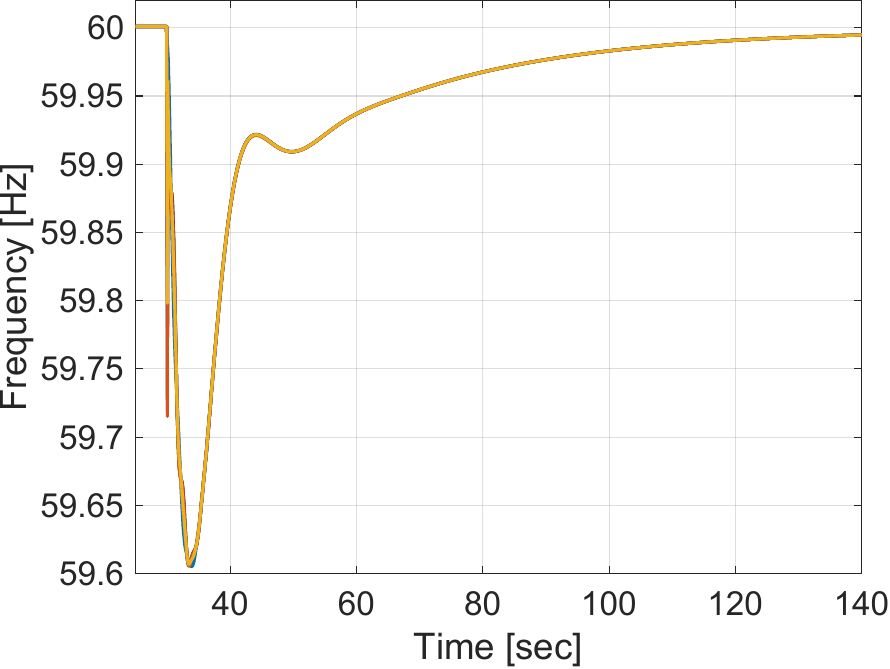}
		\vspace{-2mm}
		\caption{\texttt{RTDS} scenario I.2}        \label{fig:LFC10increaseMultipleAreas}
	\end{subfigure}
	\begin{subfigure}[t]{0.19\textwidth}
		\centering
		\includegraphics[width=0.9\linewidth]{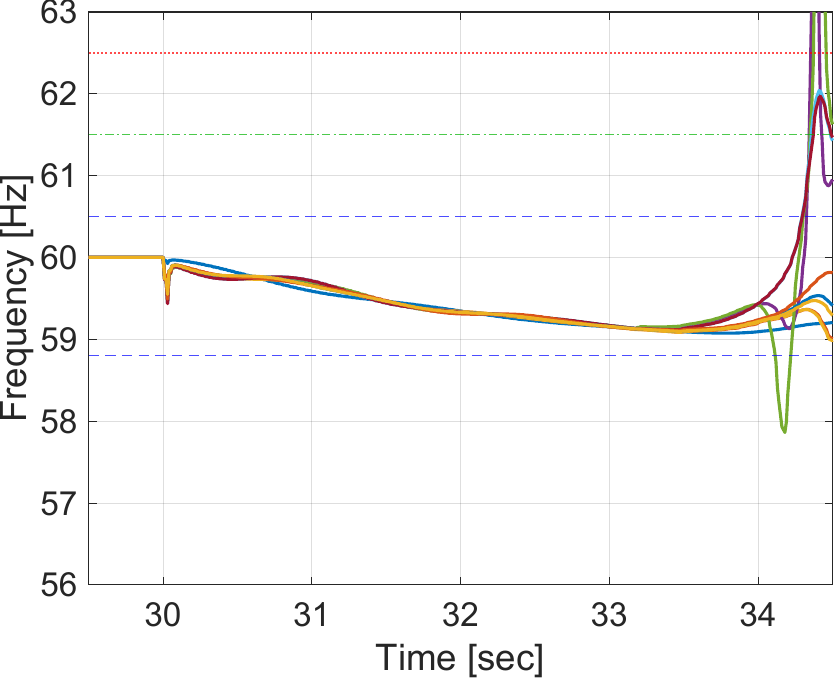}
		\vspace{-2mm}
		\caption{\texttt{RTDS} scenario I.3}        \label{fig:LFC20increase}
	\end{subfigure}
	\begin{subfigure}[t]{0.19\textwidth}
		\centering
		\includegraphics[width=0.9\linewidth]{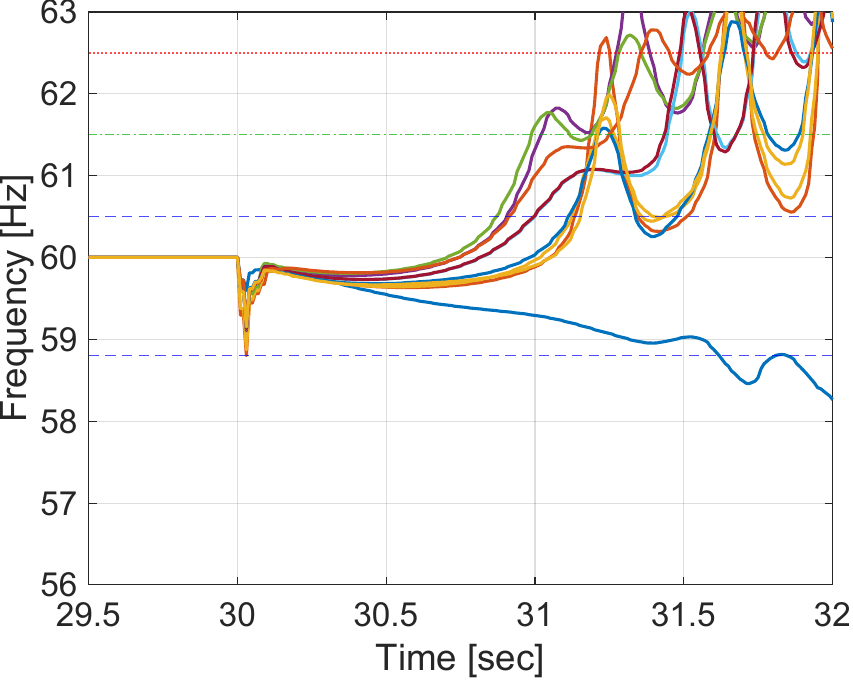}
		\vspace{-2mm}
		\caption{\texttt{RTDS} scenario I.4}        \label{fig:LFC50increase}
	\end{subfigure}
	\begin{subfigure}[t]{0.19\textwidth}
		\centering
		\includegraphics[width=0.9\linewidth]{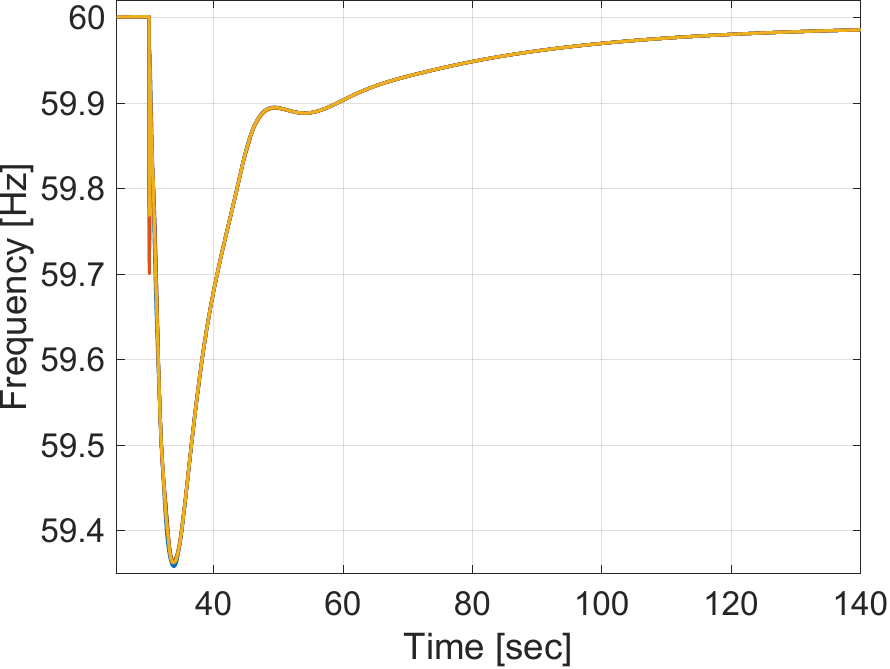}
		\vspace{-2mm}
		\caption{\texttt{RTDS} scenario I.5}
		\label{fig:LFCmax}
	\end{subfigure}    
	\vspace{-2mm}
	\caption{Frequency responses for different variants of scenario I.}
	\vspace{-3mm}
	\label{fig:ScenarioIResults}
\end{figure*}

\begin{figure}
	\begin{subfigure}[t]{0.49\linewidth}
		\centering
		\adjustbox{trim={0, 0, 0, 3mm}, clip}{%
			\def\svgwidth{\linewidth}
			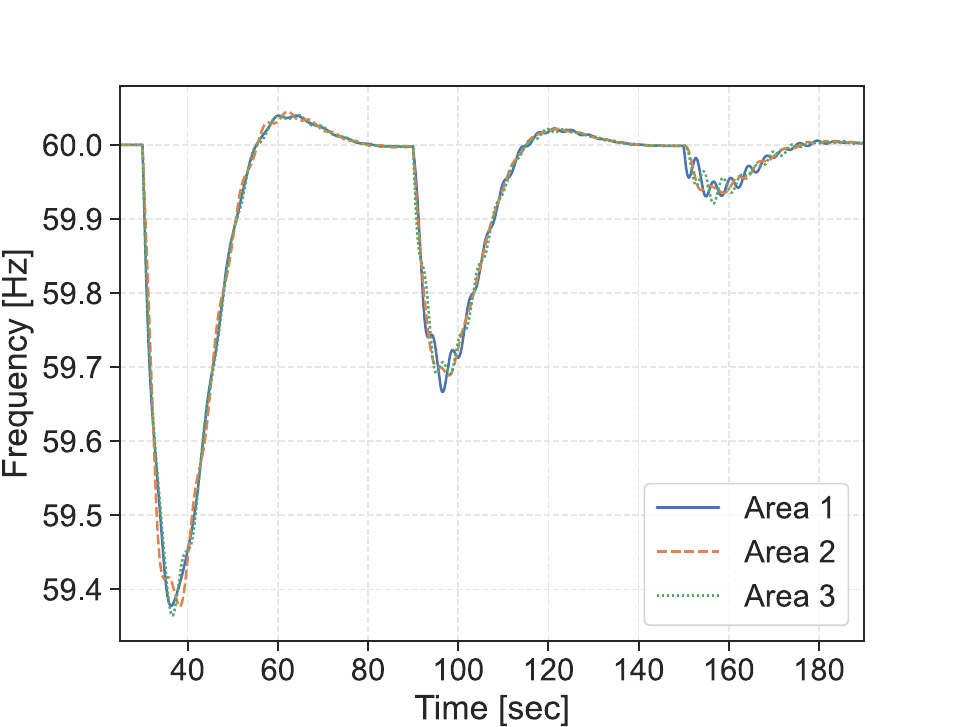}
		\vspace{-6mm}
		\caption{\texttt{Python} scenario II}
		\label{fig:pyMultistep}
	\end{subfigure}
	\begin{subfigure}[t]{0.49\linewidth}
		\centering
		\includegraphics[width=0.9\linewidth]{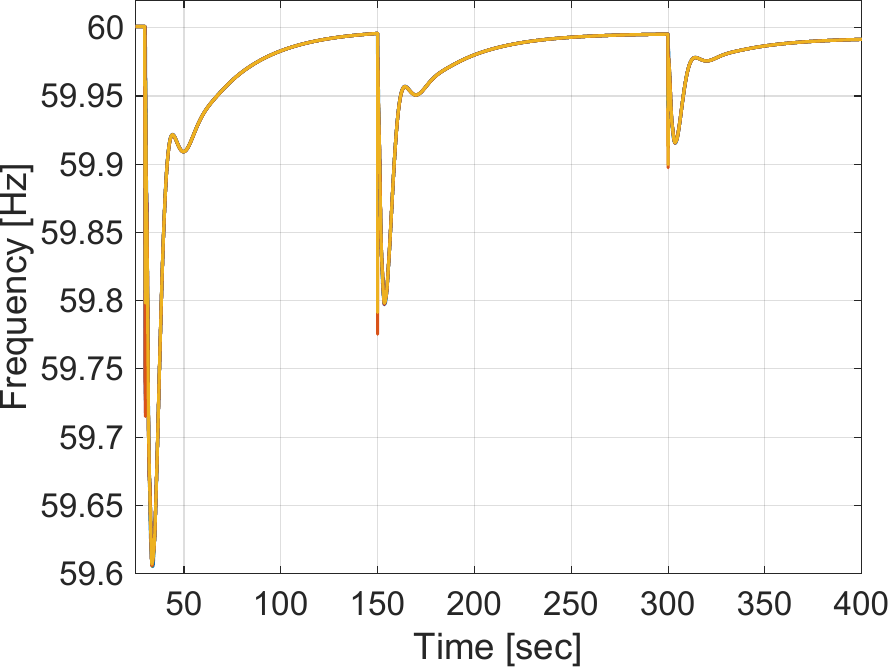}
		\vspace{-2mm}
		\caption{\texttt{RTDS} scenario II}
		\label{fig:LFCmultistep}
	\end{subfigure}
	\vspace{-2mm}
	\caption{Frequency responses for scenario II.}
	\vspace{-4mm}
	\label{fig:ScenarioIIResults}
\end{figure}

In the \texttt{Python} simulation results, the frequency plots show each area of the 39-bus system. For lower attack levels in scenarios I.1 (Fig. \ref{fig:pySimple10upArea1}) and I.2 (Fig. \ref{fig:pySimple10upAll}), the frequency drops but remains within permissible boundaries, ensuring no lasting impact on system operation. In scenario I.3 (Fig. \ref{fig:pySimple20upAll}), the frequency temporarily falls below the continuous operation range but returns to safety without exceeding the specified operational time limits. In scenario I.4 (Fig. \ref{fig:pySimple50upAll}), the system eventually restores the initial frequency. However, the frequency nadir crosses the under-frequency threshold, which could activate protection schemes. The increased volatility in Fig. \ref{fig:pySimple10upArea1} compared to Fig. \ref{fig:pySimple10upAll} results from concentrating the attack in a single area, inducing higher energy flows on tie-lines as power is redistributed from bordering areas. The attacks spread across all areas (Figs. \ref{fig:pySimple10upAll}, \ref{fig:pySimple20upAll}, and \ref{fig:pySimple50upAll}) distribute the load changes more evenly, reducing tie-line stress and leading to more synchronized frequency behavior among areas. Results for scenario I.5 (Fig. \ref{fig:pyMaxInAllAreas}) show a minimal difference from scenario II (Fig. \ref{fig:pyMultistep}), as the \texttt{Python} simulation assumes unlimited power reserves, allowing the frequency to recover to nominal values without constraints.

\begin{table}[t]
	\centering
	\caption{Saudi Arabia grid code frequency thresholds \cite{saudiGrid}.}
	\vspace{-2mm}
	\begin{tabular}{||c|c|c||}
		\hline\hline
		\textbf{\begin{tabular}{c} Below Nominal \\ Frequency [Hz]\end{tabular}} & \textbf{\begin{tabular}{c} Above Nominal \\ Frequency [Hz]\end{tabular}} & \textbf{\begin{tabular}{c} Operation \\ Requirement\end{tabular}} \\
		\hline
		58.8 - 60.0 & 60.0 - 60.5 & Continuous \\
		\hline
		57.5 - 58.7 & 60.6 - 61.5 & For 30 minutes \\
		\hline
		57.0 - 57.4 & 61.6 - 62.5 & For 30 seconds \\
		\hline\hline
	\end{tabular}
	\vspace{-5mm}
	\label{tab:SaudiThresholds}
\end{table}

\begin{figure}
	\begin{subfigure}[t]{0.49\linewidth}
		\centering
		\includegraphics[width=\linewidth]{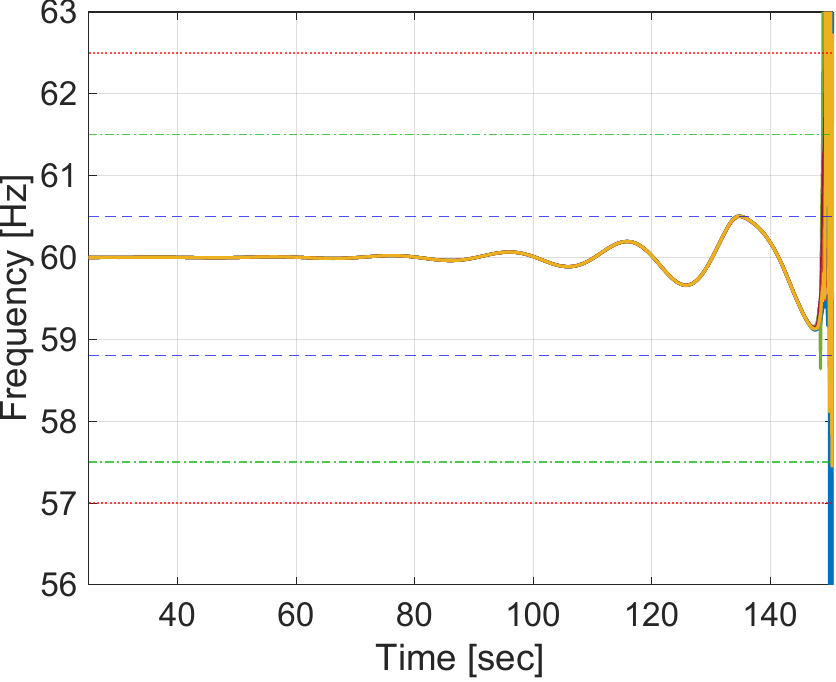}
		\vspace{-6mm}
		\caption{\texttt{RTDS} scenario III}
		\label{fig:ScenarioIIIResults}
	\end{subfigure}
	\begin{subfigure}[t]{0.49\linewidth}
		\centering
		\includegraphics[width=\linewidth]{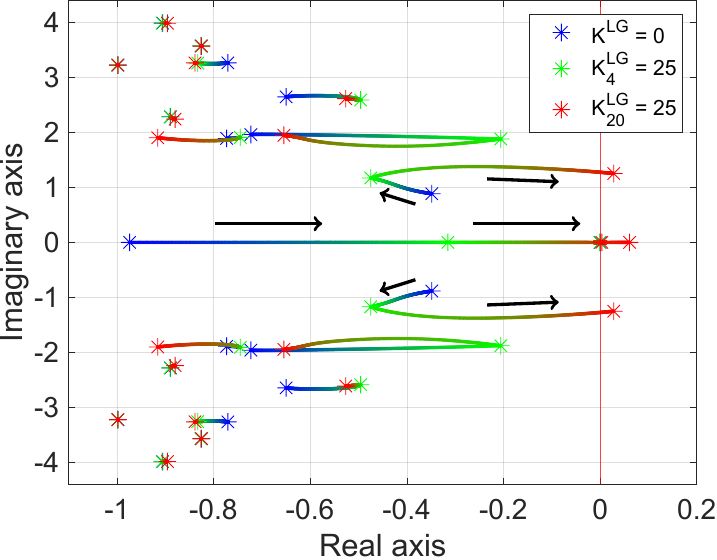}
		\vspace{-6mm}
		\caption{Eigenvalues-based stability analysis}
		\label{fig:stabilityAttack}
	\end{subfigure}
	\vspace{-1mm}
	\caption{Frequency responses and stability analysis for scenario III (DLAA).}
	\vspace{-7mm}
	\label{fig:DLAAFigure}
\end{figure}

In the \texttt{RTDS} results, the frequency plots display each of the ten generators from the 39-bus system. Quantitative data for scenarios are provided in Table \ref{tab:QuantitativeData}. The steady-state values indicate the stabilized frequency at the end of the simulation, while the settling time represents the point at which the frequency returns within acceptable limits; a ``-'' for the former metric indicates no steady state, and for the latter, the system's inability to fully recover the frequency. 
In scenario I.1 (Fig. \ref{fig:LFC10increaseSingleArea}), the system remained stable but could not fully restore nominal frequency due to limited reserves of the generators participating in LFC. In contrast, scenario I.2 (Fig. \ref{fig:LFC10increaseMultipleAreas}) led to nominal frequency restoration, as balancing smaller, distributed load changes across multiple areas proved more manageable. 
For higher load changes in scenarios I.3 (Fig. \ref{fig:LFC20increase}) and I.4 (Fig. \ref{fig:LFC50increase}), the system became unstable within seconds of LAA activation. In scenario I.5 (Fig. \ref{fig:LFCmax}), the maximum load increase the LFC could manage was 16\%. In scenario II, where the attack load increased incrementally, the system could handle a slightly higher load increase (17\%) compared to the single-step increase (16\%), as shown in Fig. \ref{fig:LFCmultistep}. This suggests that an LAA botnet uncoordinated in time may have less severe consequences than a synchronized, instantaneous load change. 
Finally, in scenario III, the corrupted load responded to minor, always present frequency fluctuations. As shown in Fig. \ref{fig:ScenarioIIIResults}, the attack's impact was initially subtle but rapidly escalated later, disrupting the system within $120$ seconds.

\begin{table*}[t]
	\centering
	\caption{Quantitative data from \texttt{RTDS} simulation by scenario number.}
	\vspace{-2mm}
	\begin{tabular}{||*{7}{c|}|}
		\hline\hline
		\textbf{Scenario} & \textbf{Steady} & \textbf{Settling} & \textbf{Frequency} & \textbf{Nadir} & \textbf{Frequency} & \textbf{Zenith} \\
		\textbf{Number} & \textbf{State [Hz]} & \textbf{Time [sec]} & \textbf{Nadir [Hz]} & \textbf{Time [sec]} & \textbf{Zenith [Hz]} & \textbf{Time [sec]} \\
		\hline
		I.1 & 59.815 & - & 59.381 & 36.00 & 60.001 & 29.73 \\
		\hline
		I.2 & 59.996 & 110.96 & 59.605 & 33.86 & 60.001 & 29.73 \\
		\hline
		I.3 & - & - & 57.863 & 34.17 & 69.293 & 34.38 \\
		\hline
		I.4 & - & - & 58.284 & 31.99 & 66.618 & 31.98 \\
		\hline
		I.5 & 59.989 & 156.57 & 59.358 & 33.81 & 60.001 & 29.73 \\
		\hline
		II & 59.992 & 357.22 & 59.605 & 33.86 & 60.001 & 29.73 \\
		\hline
		III & - & - & 58.116 & 149.98 & 76.623 & 149.18 \\
		\hline\hline
	\end{tabular}
	\vspace{-5mm}
	\label{tab:QuantitativeData}
\end{table*}

\begin{figure*}
	\begin{subfigure}[t]{0.24\textwidth}
		\centering
		\includegraphics[width=0.9\linewidth]{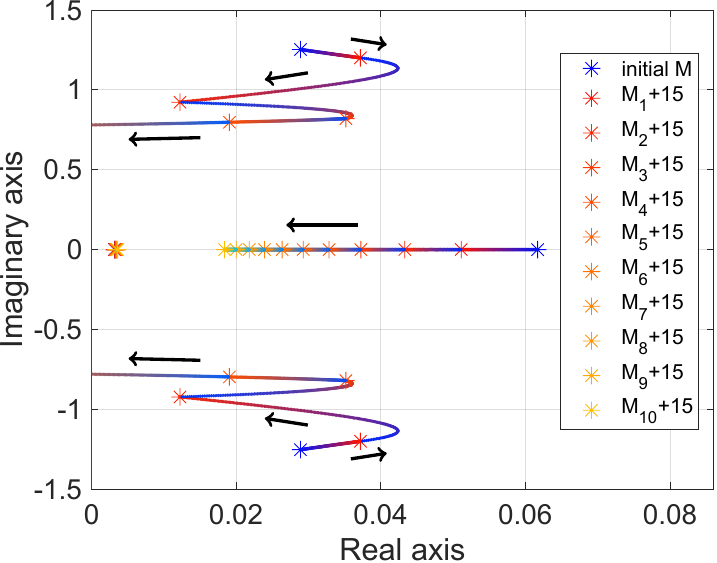}
		\vspace{-2mm}
		\caption{Each $M$ increased by 15.}
		\label{fig:stabilityM}
	\end{subfigure}
	\begin{subfigure}[t]{0.24\textwidth}
		\centering
		\includegraphics[width=0.9\linewidth]{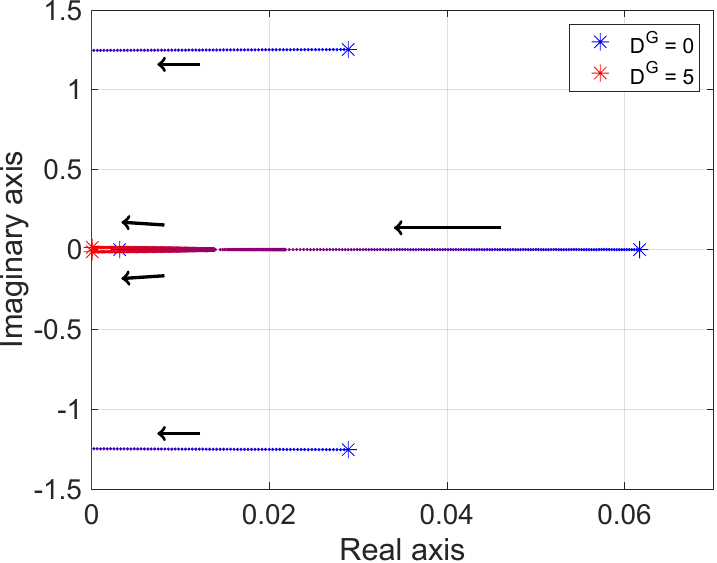}
		\vspace{-2mm}
		\caption{$D^G_2$ increased to 5.}
		\label{fig:stabilityD_GSuccess}
	\end{subfigure}
	\begin{subfigure}[t]{0.24\textwidth}
		\centering
		\includegraphics[width=0.9\linewidth]{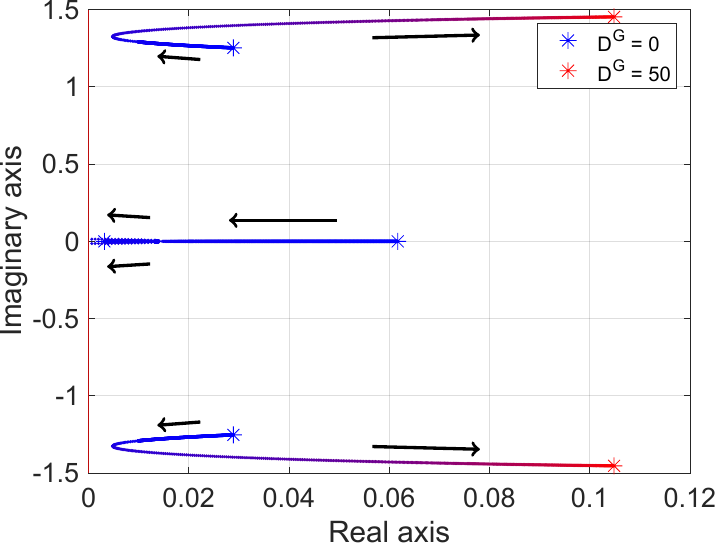}
		\vspace{-2mm}
		\caption{$D^G_6$ increased to 50}
		\label{fig:stabilityD_GFailure}
	\end{subfigure}
	\begin{subfigure}[t]{0.24\textwidth}
		\centering
		\includegraphics[width=0.9\linewidth]{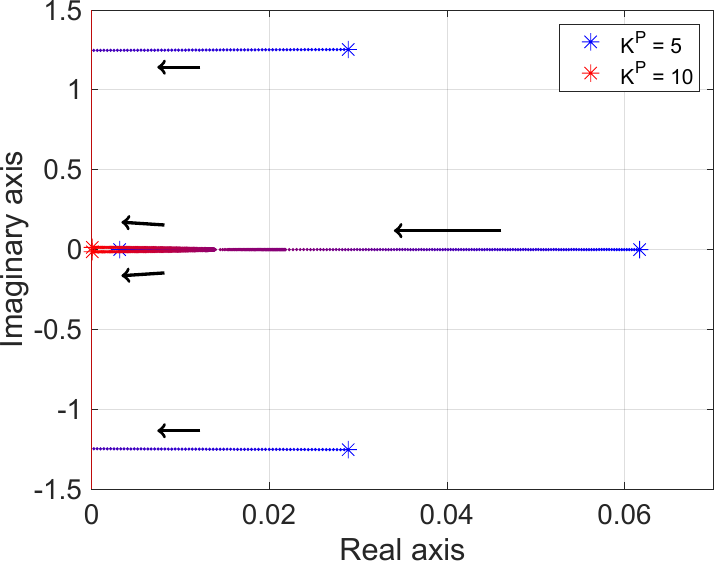}
		\vspace{-2mm}
		\caption{$K^P_2$ increased to 10.}
		\label{fig:stabilityK_Psuccess}
	\end{subfigure}
	
	\hfil
	
	\begin{subfigure}[t]{0.24\textwidth}
		\centering
		\includegraphics[width=0.9\linewidth]{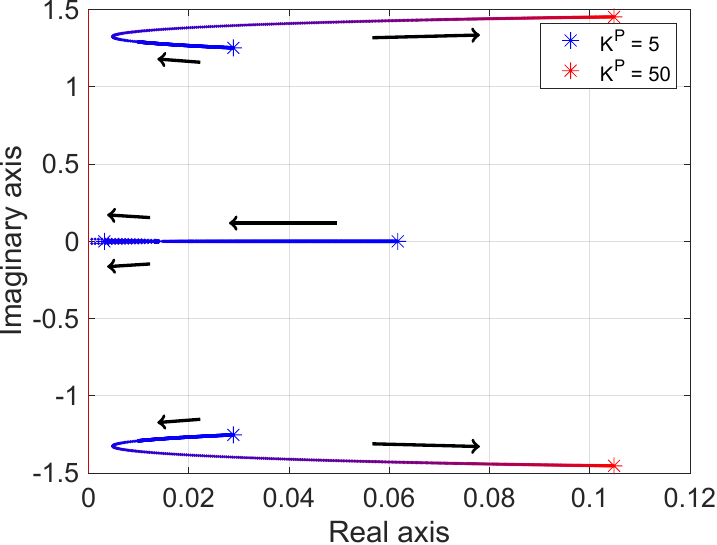}
		\vspace{-2mm}
		\caption{$K^P_6$ increased to 50.}
		\label{fig:stabilityK_Pfailure}
	\end{subfigure}
	\begin{subfigure}[t]{0.24\textwidth}
		\centering
		\includegraphics[width=0.9\linewidth]{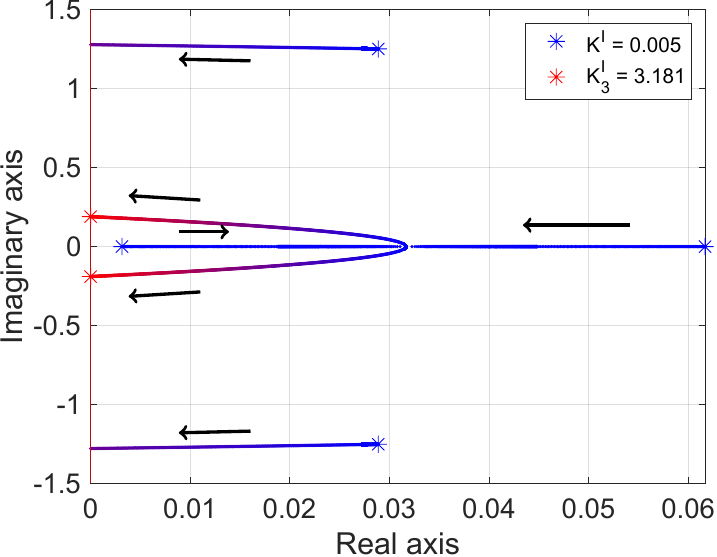}
		\vspace{-2mm}
		\caption{$K^I_3$ increased to 3.181.}
		\label{fig:stabilityK_I3}
	\end{subfigure}
	\begin{subfigure}[t]{0.24\textwidth}
		\centering
		\includegraphics[width=0.9\linewidth]{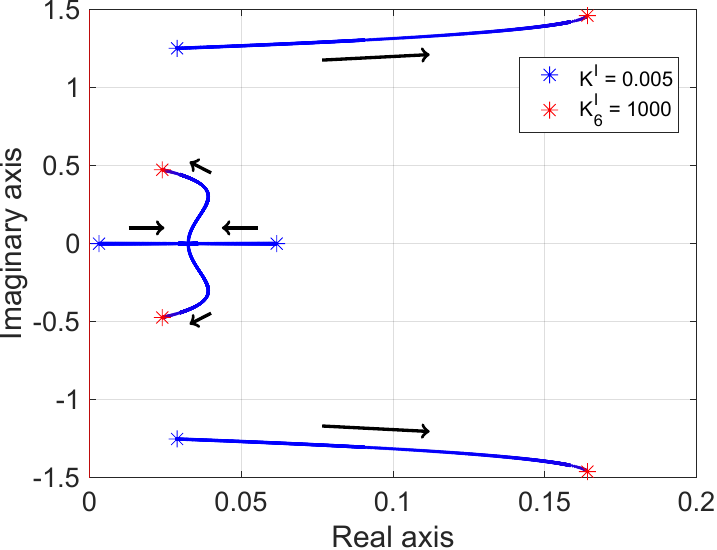}
		\vspace{-2mm}
		\caption{$K^I_6$ increased to 1000.}
		\label{fig:stabilityK_I6}
	\end{subfigure}
	\begin{subfigure}[t]{0.24\textwidth}
		\centering
		\includegraphics[width=0.9\linewidth]{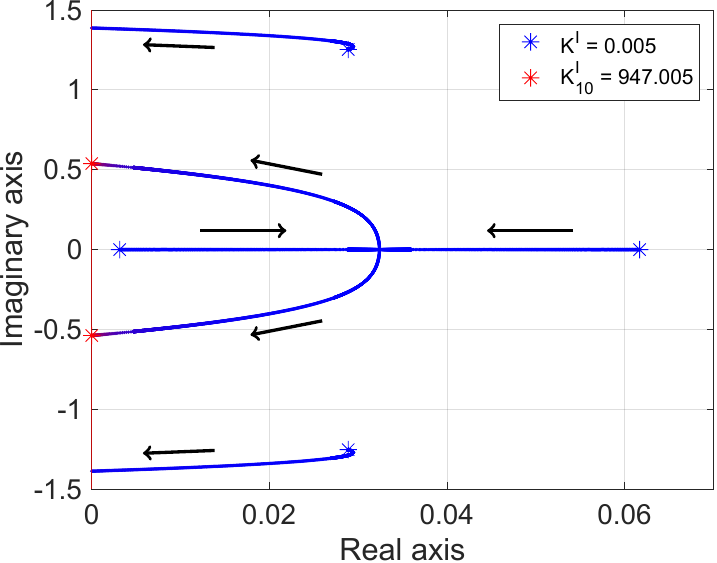}
		\vspace{-2mm}
		\caption{$K^I_{10}$ increased to 947.005.}
		\label{fig:stabilityK_I10}
	\end{subfigure}
	\vspace{-1mm}
	\caption{Eigenvalues-based analysis of the effects of different parameters adjustments on the system stability under the DLAA.}
	\label{fig:stabilityParameters}
	\vspace{-5mm}
\end{figure*}

\vspace{-1mm}
\subsection{Comparison of Simulation Results}
\label{subsec:ComparisonOfSimulationResults}
\vspace{-1mm}

The analysis of simulation results reveals the accuracy of \texttt{Python} simulation in modeling the theoretical behavior of LFC under LAA. However, it fails to account for physical limitations, such as power reserve constraints. In contrast, the \texttt{RTDS} simulation captures the nonlinearities of LFC and limitations of system components, such as valve position and generation limits. These distinctions enabled the capture of system behaviors under LAA that the \texttt{Python} simulation could not differentiate. 
There, regardless of attack magnitude, the steady state always returned to nominal frequency, with disruption removal occurring at a consistent rate following the attack. However, in \texttt{RTDS} scenarios I.2, I.5, and II, recovery times were considerably slower, with a maximum difference of $167$ seconds for scenario II. Additionally, in \texttt{Python} scenario I.4, the frequency drop below the lowest operational threshold suggested potential equipment damage or disconnection in practice. In \texttt{RTDS} scenarios I.3 and I.4, significant frequency instability rendered the system unrecoverable. 
The \texttt{Python} scenarios I.5 and II produced similar steady states. However, the higher fidelity of the \texttt{RTDS} simulation revealed that multistep attacks of equivalent magnitude require further load increase compared to a single-step increase. All these findings suggest that the \texttt{RTDS} simulation provides a more accurate representation of LAA impacts on LFC.

\vspace{-1mm}
\subsection{System Stability Assessment}
\label{subsec:SystemStabilityAssessment}
\vspace{-1mm}

To further explore scenario III, we present the eigenvalues plot of $A^*$ \eqref{eq:matrixA*Dynamic} for different values of $K^{LG}$. Then, we investigate DLAA countermeasures by adjusting system parameters to restore stability. Similarly to the \texttt{RTDS} simulation, we include the LFC only at generators 3, 6, and 10. For others, the value of $K^I$ equals zero. Each plot shows the eigenvalues' progression under the changing parameter values. The initial positions of eigenvalues for the parameter's original value are marked with the blue "$\Asterisk$". Fig. \ref{fig:stabilityAttack} illustrates the impact of DLAA by consecutively increasing $K^{LG}$ at two different buses. Combined, these two changes move four eigenvalues to the positive real plane, rendering the DLAA successful. In contrast, in Fig. \ref{fig:stabilityParameters}, each scenario initially includes DLAA effects from Fig. \ref{fig:stabilityAttack} and investigates the attack countermeasures.

To protect the system, we consider changes to its parameters: $M$, $D^G$, $K^P$, and $K^I$. We increase each parameter's value for different generators to negate the attack's effects. In Fig. \ref{fig:stabilityM}, we consecutively increase $M$ by 15 at each generator. It results in two of four eigenvalues returning to the negative plane. However, with the other two remaining positive, the system stays unstable. Next, we separately increase $D^G$ by 5 at generators 2 to 5 and 10. For generator 1, the increase of 8.33 resulted in a similar eigenvalues movement. In both cases, all eigenvalues became negative, preventing the DLAA. In Fig. \ref{fig:stabilityD_GSuccess}, we present results only for generator two, as they are analogous to other cases. For generators 6 to 9, even after raising the parameter value to 50, as shown in Fig \ref{fig:stabilityD_GFailure} for generator 6, the system remained unstable. Parameters $M$ and $D^G$ are derived from the generator's physical properties \cite{RobustPowerSystemFrequencyControl}, making them difficult to adjust. However, the adversary can readily adapt the $K^{LG}$ value for the targeted system \cite{DynamicLAA}.

In contrast, the primary and secondary control can be conveniently reconfigured to match the changing conditions \cite{RobustPowerSystemFrequencyControl}. For the $K^P$, we increased its value by 5 separately for generators 2 to 5 and 10, and also by 8.33 for generator 1. In both cases, this single change restored the system stability, as shown for generator 2 in Fig. \ref{fig:stabilityK_Psuccess}. However, despite increasing $K^P$ up to 50 at generators 6 to 9, as shown for generator 6 in Fig. \ref{fig:stabilityK_Pfailure}, the system remained unstable. Lastly, we investigated the parameter $K^I$. To achieve stability, we increased its value to 3.181 at generator 3 (Fig. \ref{fig:stabilityK_I3}). For generator 6, despite increasing the value to 1000, as shown in Fig. \ref{fig:stabilityK_I6}, the system remained unstable. Finally, for generator 10, the attack is prevented when the $K^I$ equals 947.005, as shown in Fig. \ref{fig:stabilityK_I10}. However, such a high LFC coefficient value seems unsuitable for practical solutions. 

\vspace{-2mm}
\section{Conclusion}
\label{sec:Conclusion}
\vspace{-1mm}

This paper uses analytical models and simulations to analyze the link between LAA and LFC. A \texttt{Python}-based simulation investigates unconstrained system responses to SLAA and Multistep SLAA, while an \texttt{RTDS}-based simulation examines transient responses to SLAAs and DLAAs, detailing component behavior and limitations. Eigenvalue-based stability analysis in \texttt{MATLAB} explores parameter adjustments for frequency stability under DLAA. Results highlight LAA's impact on LFC and demonstrate approaches to disrupt the system. Stability analysis depicts how parameter tuning, particularly in primary and secondary control coefficients, can mitigate DLAAs.

\vspace{-2mm}
\section*{Acknowledgments}
\vspace{-1mm}
This publication is based upon
work supported by King Abdullah University of Science and Technology under Award No. ORFS-2022-CRG11-5021.

\vspace{-2mm}
\bibliographystyle{ieeetr}
\bibliography{references}

\vfill

\end{document}